\documentclass[reprint,aps,prd,nofootinbib,superscriptaddress]{revtex4-1}
\usepackage{amsmath,amssymb}
\usepackage{graphicx}
\usepackage{booktabs}
\usepackage{xcolor}

\graphicspath{{figures/}}

\begin{document}

\title{Hawking Radiation from the Dymnikova Regular Black Hole}

\author{Milena Skvortsova}
\email{milenas577@mail.ru}
\affiliation{Peoples' Friendship University of Russia (RUDN University), 6 Miklukho-Maklaya Street, Moscow, 117198, Russia}

\date{\today}

\begin{abstract}
We study Hawking radiation of the Dymnikova regular black hole.  This model replaces the central singularity by a smooth de Sitter core while remaining Schwarzschild-like far from the hole, and its black-hole branch ends in a cold extremal remnant.  We compute the greybody factors of the Standard Model test fields and gravitons, compare the precise numerical scattering results with WKB estimates, and use the resulting spectra to estimate an adiabatic evaporation history.  The main effect is not a dramatic change in the transmission probabilities: the greybody thresholds move only slightly as the geometry approaches the remnant.  Instead, the rapid decrease of the Hawking temperature strongly suppresses the total luminosity.  The photon, light-fermion and graviton channels all fade near the endpoint, with the gravitational contribution remaining subdominant.  The residual massless flux becomes increasingly fermion dominated because the photon channel is suppressed more efficiently.  The black hole approaches the cold remnant only asymptotically, so the quoted lifetime estimates should be interpreted as cutoff times to near-extremal configurations rather than as complete evaporation times.  
\end{abstract}

\maketitle

\section{Introduction}

Hawking radiation is thermal at the event horizon, but it is not observed at infinity as an ideal black-body spectrum.  Each outgoing field mode must propagate through the curvature potential outside the horizon.  This scattering problem produces a frequency-dependent transmission probability, the greybody factor, which multiplies the Bose--Einstein or Fermi--Dirac occupation factor and determines the absorption probability, partial-wave cross section and luminosity of the black hole \cite{Hawking1975,Page1976,Unruh1976,DasGibbonsMathur1997}.  Greybody factors are therefore an important probe of both the near-horizon temperature and the full exterior geometry.

Regular black holes replace the Schwarzschild singularity by a finite core while keeping an asymptotically flat exterior.  This makes them natural phenomenological laboratories for testing how short-distance modifications alter scattering and evaporation.  Various radiation phenomena around regular black holes have therefore been analyzed in a broad range of studies, including greybody filtering, Hawking spectra, quasinormal ringing, absorption cross sections, shadow-related signatures and late-time evaporation scenarios \cite{Konoplya:2025ect,Skvortsova:2024wly,Li:2014fka,Dubinsky:2026wcv,
Yang:2021cvh,Held:2019xde,Bolokhov:2025fto,Pedraza:2021hzw,
Mukohyama:2023xyf,Skvortsova:2025cah,Konoplya:2023ppx,
MahdavianYekta:2019pol,Cai:2021ele,Guo:2024jhg,
Bolokhov:2026eqf,Lopez:2022uie,Huang:2023aet,
Fernando:2012yw,Konoplya:2022hll,Lin:2013ofa,
Jusufi:2020odz,DuttaRoy:2022ytr,SalehThomasKofane2018,
Konoplya:2026gim,Al-Badawi:2023lke,RinconSantos2020,
Jawad:2020hju,Gingrich:2024tuf,Skvortsova:2026unq,
Konoplya:2023ahd,Zhang:2024nny,Bolokhov:2023ruj,
Flachi:2012nv,Konoplya:2023aph}.  The Dymnikova solution is one of the earliest and simplest examples \cite{Dymnikova:1992ux}.  Its mass function rises exponentially from the origin, giving a de Sitter core and an exterior that approaches Schwarzschild at large radius.  Because the solution is neutral, its test-field radiation can be compared directly with the standard uncharged Page benchmarks without electric Schwinger-pair-production complications.

Dymnikova-type metrics are also useful beyond their original phenomenological motivation.  Exponential or regularized mass functions appear in discussions of renormalization-group-improved and asymptotically safe black holes, where running gravitational couplings or effective short-distance cutoffs soften the central region \cite{Bonanno:2000ep,Konoplya:2023aph,Lutfuoglu:2025ohb}.  The precise microscopic interpretation differs from model to model, but the scattering question is common: once a regular core and a possible extremal remnant are present, do the greybody factors enhance the outgoing flux, or does the cooling of the horizon dominate?

Here we answer this question for electromagnetic and massless Dirac test fields and for gravitational perturbations on the Dymnikova background.  The spin-1 and spin-1/2 sectors are the relevant massless sectors for a photon-plus-fermion Page-type comparison and for a simple Standard-Model-inspired multiplicity count, while the spin-2 sector is included through an axial effective potential.  Because only the axial equation is solved explicitly, the gravitational luminosity below assumes that the polar and axial greybody factors are approximately the same.  We keep the ADM mass fixed to $M=1$, determine the extremal value of the remaining length parameter, compute greybody factors by direct integration of the radial wave equation, compare the barrier-top behavior with first- and third-order WKB estimates, integrate the Hawking spectra including the gravitational channel, and estimate the adiabatic evaporation time for a fixed Dymnikova core scale.

The calculation is semiclassical and should be read in the regime where the instantaneous black-hole mass is much larger than the energy of a typical emitted quantum, $\omega/M\ll1$.  If the core scale is tied directly to the Planck length, the final endpoint is expected to require genuine quantum-gravity input; the present test-field treatment is safest either for $M\gg M_{\rm Pl}$ before the last stage or for an effective regularization scale $r_0$ larger than the Planck length.  In this regime individual Standard-Model quanta are much lighter than the black hole, so recoil from a single emission is negligible.  The separate question of whether a massive species is thermally populated is controlled by its rest mass relative to $T_H$; consequently the massless Standard-Model-inspired count used below is a proxy, while massive thresholds are left for future work.

The main result is selective quenching.  The electromagnetic, Dirac and gravitational greybody factors change only mildly along most of the branch, and the Dirac thresholds even move slightly downward close to extremality.  The temperature, however, falls much more rapidly and vanishes at the double horizon.  Consequently the energy-emission rates collapse as the remnant is approached.  The photon channel is suppressed more efficiently than the Dirac channel, so a massless Standard-Model-like flux becomes progressively fermion dominated.  Within the axial--polar equality assumption, the gravitational channel follows the same thermal suppression and remains numerically smaller than the photon-plus-light-fermion proxy.

\section{Dymnikova Geometry and Extremality}

The metric considered here is the static, spherically symmetric regular black-hole geometry introduced by Dymnikova as a nonsingular vacuum configuration with a de Sitter core \cite{Dymnikova:1992ux,Dymnikova:2004qg}.  It also appears naturally in asymptotically safe gravity (ASG): a self-consistent iterative RG improvement of Schwarzschild spacetime was shown to converge, when the high-energy gravitational flow reaches a non-Gaussian fixed point, to a singularity-free Dymnikova-type geometry \cite{Platania:2019kyx}.  In the notation used below, the line element is
\begin{equation}
\label{eq:metric}
 ds^2=-f(r)dt^2+\frac{dr^2}{f(r)}+r^2d\Omega_2^2,
\end{equation}
with
\begin{equation}
\label{eq:dymnikova_original}
 f(r)=1-\frac{2M}{r}\left[1-\exp\!\left(-\frac{r^3}{2Mr_0^2}\right)\right].
\end{equation}
Here $M$ is the ADM mass and $r_0$ is the length scale that controls the central de Sitter core.  Throughout the paper we use natural units $G=c=\hbar=k_B=1$, so masses, lengths, frequencies and temperatures are all expressed in powers of one length scale.  The symbol $h$ is only a dimensionless deformation parameter after setting $M=1$; it should not be confused with Planck's constant.  In units $M=1$ it is convenient to define
\begin{equation}
\label{eq:h_definition}
 h^3=2r_0^2,
 \qquad
 f(r)=1-\frac{2}{r}\left[1-\exp\!\left(-\left(\frac{r}{h}\right)^3\right)\right].
\end{equation}
The Schwarzschild limit is recovered for $h\to0$, or more practically for $r/h\gg1$ outside the horizon.  Near the origin,
\begin{equation}
 f(r)=1-\frac{2r^2}{h^3}+O(r^5),
\end{equation}
so the core is de Sitter with effective cosmological constant $\Lambda_{\rm eff}=6/h^3$.

The parameter $r_0$ is held fixed only in the evaporation estimate of Sec.~\ref{sec:lifetime}.  In the preceding fixed-mass scattering calculation, changing $h$ means comparing different members of the one-parameter family at the same ADM mass.  The outer horizon $r_+$ is the largest positive root of $f(r)=0$.  The Hawking temperature is
\begin{equation}
\label{eq:temperature}
 T_H=\frac{f'(r_+)}{4\pi}.
\end{equation}
The extremal endpoint is obtained from
\begin{equation}
\label{eq:extremal_conditions}
 f(r_{\rm ext})=0,
 \qquad
 f'(r_{\rm ext})=0.
\end{equation}
Introducing $u=(r/h)^3$, the two conditions reduce to
\begin{equation}
\label{eq:u_ext}
 e^u=1+3u,
 \qquad
 r_{\rm ext}=2(1-e^{-u}),
 \qquad
 h_{\rm ext}=\frac{r_{\rm ext}}{u^{1/3}}.
\end{equation}
The nonzero solution is
\begin{equation}
\label{eq:extremal_numbers}
\begin{aligned}
 u_{\rm ext}&=1.903813694440,\\
 r_{\rm ext}&=1.702001406975,\\
 h_{\rm ext}&=1.373256824469.
\end{aligned}
\end{equation}
Equivalently, in the original convention of Eq.~\eqref{eq:dymnikova_original},
\begin{equation}
 r_{0,{\rm ext}}=\sqrt{\frac{h_{\rm ext}^3}{2}}=1.137922411780.
\end{equation}
For $0\le h<h_{\rm ext}$ the black hole has a nonzero temperature; at $h=h_{\rm ext}$ the horizons merge and $T_H=0$.

\begin{figure*}[t]
\centering
\includegraphics[width=0.92\textwidth]{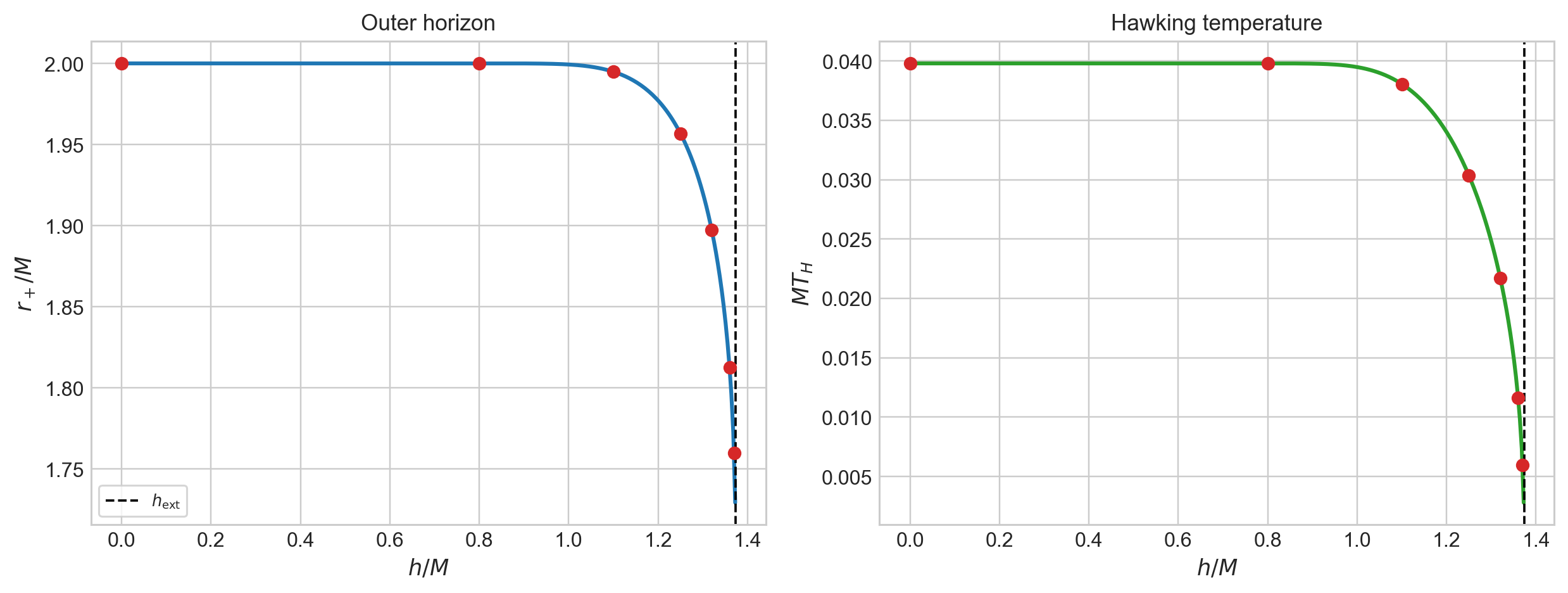}
\caption{Outer horizon radius and Hawking temperature for the Dymnikova black hole in units $M=1$.  The dashed vertical line marks $h_{\rm ext}=1.373256824469$.  The temperature vanishes at the extremal double horizon.}
\label{fig:geometry}
\end{figure*}

Table~\ref{tab:geometry} lists the benchmark values used in the greybody calculation.  Figure~\ref{fig:geometry} shows the corresponding horizon radius and temperature.  The temperature remains close to the Schwarzschild value $1/(8\pi)$ until the deformation scale becomes comparable to its extremal value, and then falls rapidly.

\begin{table}[t]
\centering
\caption{Dymnikova benchmark geometries in units $M=1$.  The final column gives the fraction of the extremal parameter value.}
\label{tab:geometry}
\begin{ruledtabular}
\begin{tabular}{cccc}
$h$ & $r_+$ & $T_H$ & $h/h_{\rm ext}$ \\
        0.0 & 2.000000 & 0.039789 & 0.000000 \\
        0.8 & 2.000000 & 0.039788 & 0.582557 \\
        1.1 & 1.994862 & 0.038053 & 0.801015 \\
        1.25 & 1.956870 & 0.030349 & 0.910245 \\
        1.32 & 1.897385 & 0.021731 & 0.961219 \\
        1.36 & 1.812486 & 0.011650 & 0.990346 \\
        1.37 & 1.759880 & 0.005984 & 0.997628 \\
\end{tabular}
\end{ruledtabular}
\end{table}

\section{Field Equations and Numerical Method}

For a static spherical metric of the form \eqref{eq:metric}, massless test fields reduce to one-dimensional wave equations in the tortoise coordinate
\begin{equation}
\label{eq:tortoise}
 \frac{dr_*}{dr}=\frac{1}{f(r)}.
\end{equation}
The master equation is
\begin{equation}
\label{eq:master}
 \frac{d^2\Psi}{dr_*^2}+\left[\omega^2-V_s(r)\right]\Psi=0.
\end{equation}
For electromagnetic perturbations the potential is \cite{Konoplya:2011qq} 
\begin{equation}
\label{eq:em_potential}
 V_{\ell}^{\rm EM}(r)=f(r)\frac{\ell(\ell+1)}{r^2},
 \qquad \ell=1,2,\ldots .
\end{equation}
For a massless Dirac field we use the usual supersymmetric partner form.  The two chiral partner potentials have the same transmission probability, and we take (see, for example, \cite{Konoplya:2017tvu,Cho:2004wj,Cho:2003qe,Konoplya:2007zx,Kanti:2006ua} for details)
\begin{equation}
\label{eq:dirac_potential}
\begin{aligned}
 V_k^{\rm D}(r)&=W^2+\frac{dW}{dr_*},\\
 W(r)&=\frac{k\sqrt{f(r)}}{r},
 \qquad k=1,2,\ldots .
\end{aligned}
\end{equation}
For gravitational perturbations we use the axial effective potential for Dymnikova black holes given in Ref.~\cite{Dubinsky:2025DymnikovaGravGBF}.  In the notation of Eq.~\eqref{eq:metric}, it reads
\begin{equation}
\label{eq:grav_potential}
\begin{aligned}
  V_{\ell}^{\rm grav,ax}(r)
  &=f(r)\left[\frac{(\ell+2)(\ell-1)+2f(r)}{r^2}
  -\frac{f'(r)}{r}\right],\\
  &\hspace{2.6cm}\ell=2,3,\ldots .
\end{aligned}
\end{equation}
This form is obtained by reducing the odd-parity metric perturbations to a Regge--Wheeler-type master equation on the effective Dymnikova background.  The derivation treats the regular-black-hole source as the fixed effective matter sector supporting the geometry and does not solve independent pert,urbations of that anisotropic source (see details in \cite{Ashtekar:2018cay,Bouhmadi-Lopez:2020oia,Konoplya:2025hgp,Konoplya:2024lch,Bolokhov:2025egl,Lutfuoglu:2026rqe,Lutfuoglu:2025pzi}).  It also does not provide a separate polar master potential.  We therefore use Eq.~\eqref{eq:grav_potential} for the directly computed spin-2 greybody factors and make the additional working assumption
\begin{equation}
\label{eq:axial_polar_assumption}
 \Gamma_{\ell}^{\rm grav,pol}(\omega)\simeq \Gamma_{\ell}^{\rm grav,ax}(\omega)
\end{equation}
when constructing the total graviton emission rate.  For $f=1-2/r$, Eq.~\eqref{eq:grav_potential} reduces to the standard Schwarzschild Regge--Wheeler potential $V_\ell=f[\ell(\ell+1)/r^2-6/r^3]$, where axial and polar channels are isospectral.
The multiplicity of a photon partial wave is $2(2\ell+1)$, while the multiplicity of one two-helicity massless spin-$1/2$ species is $2k$ in the convention used here.  For gravitational emission we use $2(2\ell+1)$, with the factor of two implementing the approximate axial-plus-polar degeneracy in Eq.~\eqref{eq:axial_polar_assumption}.

The boundary conditions for the greybody problem are purely ingoing at the horizon and a superposition of incoming and outgoing waves at infinity:
\begin{equation}
\Psi\sim e^{-i\omega r_*},\qquad r\to r_+,
\end{equation}
\begin{equation}
\Psi\sim A_{\rm in}e^{-i\omega r_*}+A_{\rm out}e^{+i\omega r_*},
\qquad r\to\infty .
\end{equation}
Direct integration outward from a near-horizon point gives
\begin{equation}
\label{eq:gamma_direct}
 \Gamma(\omega)=\frac{1}{|A_{\rm in}|^2}.
\end{equation}
The numerical extraction was performed on a dense $r$ grid, with the tortoise coordinate constructed by cumulative quadrature.  At large radius the final segment of the numerical solution was projected onto the two plane-wave components to determine $A_{\rm in}$ and $A_{\rm out}$ \cite{Tan:2026itp,Tan:2026vif,Page1976}.  The maximum flux-balance residual, $|\Gamma+|A_{\rm out}/A_{\rm in}|^2-1|$, over all tabulated curves was $3.73\times10^{-4}$.

We also computed first- and third-order WKB estimates.  At first order,
\begin{equation}
\label{eq:wkb_first}
 \Gamma_{\rm WKB}^{(1)}=\left[1+\exp\left(\frac{2\pi(V_0-\omega^2)}{\sqrt{-2V_0''}}\right)\right]^{-1},
\end{equation}
where $V_0$ is the barrier maximum and primes denote derivatives with respect to $r_*$.  The third-order calculation follows the Iyer--Will continuation formula, retaining the correction terms through $\Lambda_3$ and using derivatives through order six at the potential maximum \cite{SchutzWill1985,IyerWill1987,Konoplya2003WKB,KonoplyaZhidenkoZinhailo2019,Konoplya:2026rjh}.WKB methods at different orders have been used extensively in black-hole scattering and quasinormal-mode calculations, with numerous applications showing good agreement with direct numerical results for sufficiently large multipole number and near the top of a single smooth potential barrier \cite{
Bolokhov:2026uol,Momennia:2018hsm,Lutfuoglu:2025hjy,Konoplya:2010vz,Bolokhov:2024bke,Fernando:2016ftj,Skvortsova:2026jtx,Bolokhov:2023dxq,Konoplya:2024kih,Guo:2020caw,Lutfuoglu:2026pgn,
Konoplya:2009hv,Bolokhov:2026dzn,Kodama:2009bf,Eniceicu:2019npi,Malik:2026lfj,Konoplya:2019ppy,Skvortsova:2024eqi,Bolokhov:2026kqu,Konoplya:2023moy,Breton:2017hwe,Bolokhov:2024ixe,Lutfuoglu:2025ljm,Konoplya:2024hfg,Wongjun:2019ydo,Bolokhov:2026dfg,Konoplya:2010kv,Bolokhov:2025lnt,Karmakar:2023cwg,
Skvortsova:2026idf,Malik:2025erb,Bolokhov:2023bwm,Lutfuoglu:2025ljm}.  For the exponential Dymnikova metric these derivatives were evaluated by a local polynomial fit in $r_*$ around the barrier maximum.  This method is accurate near the top of a single smooth barrier, but it is not expected to be uniformly reliable deep in the tunneling tail or for the lowest Dirac modes, where the asymptotic correction remains sensitive to the local derivative fit.

\begin{figure*}[t]
\centering
\includegraphics[width=0.94\textwidth]{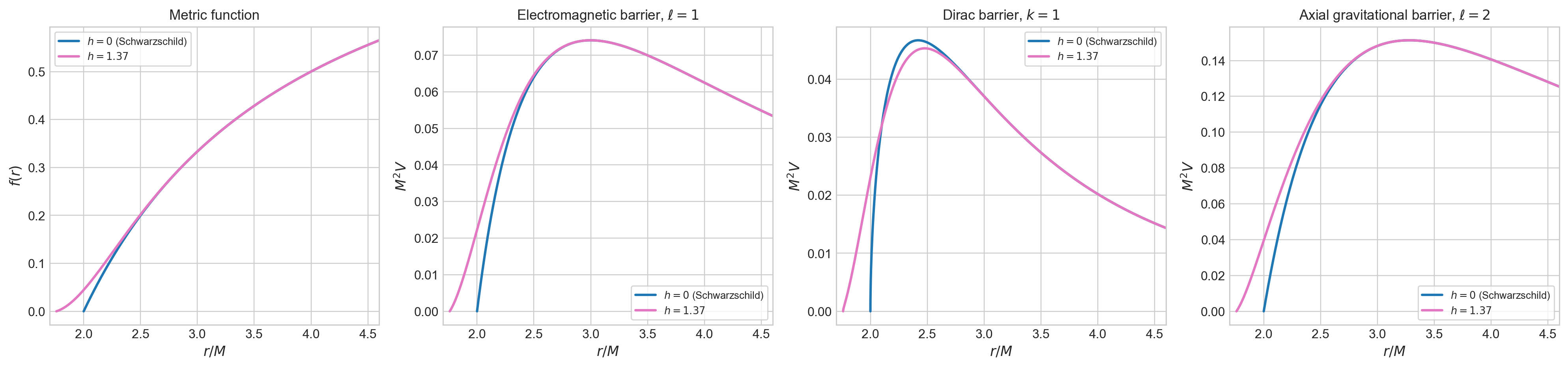}
\caption{Metric function and representative effective potentials for the Schwarzschild case and the near-extremal Dymnikova geometry.  The radial window is restricted to the barrier region to make the difference between $h=0$ and $h=1.37$ visible, and the horizon limits $f(r_+)=V(r_+)=0$ are included explicitly.  The electromagnetic barrier shown is $\ell=1$, the Dirac barrier shown is $k=1$, and the gravitational barrier shown is the axial $\ell=2$ potential of Eq.~\eqref{eq:grav_potential}.}
\label{fig:potentials}
\end{figure*}

\section{Greybody Factors and WKB Comparison}

\begin{figure*}[t]
\centering
\includegraphics[width=0.94\textwidth]{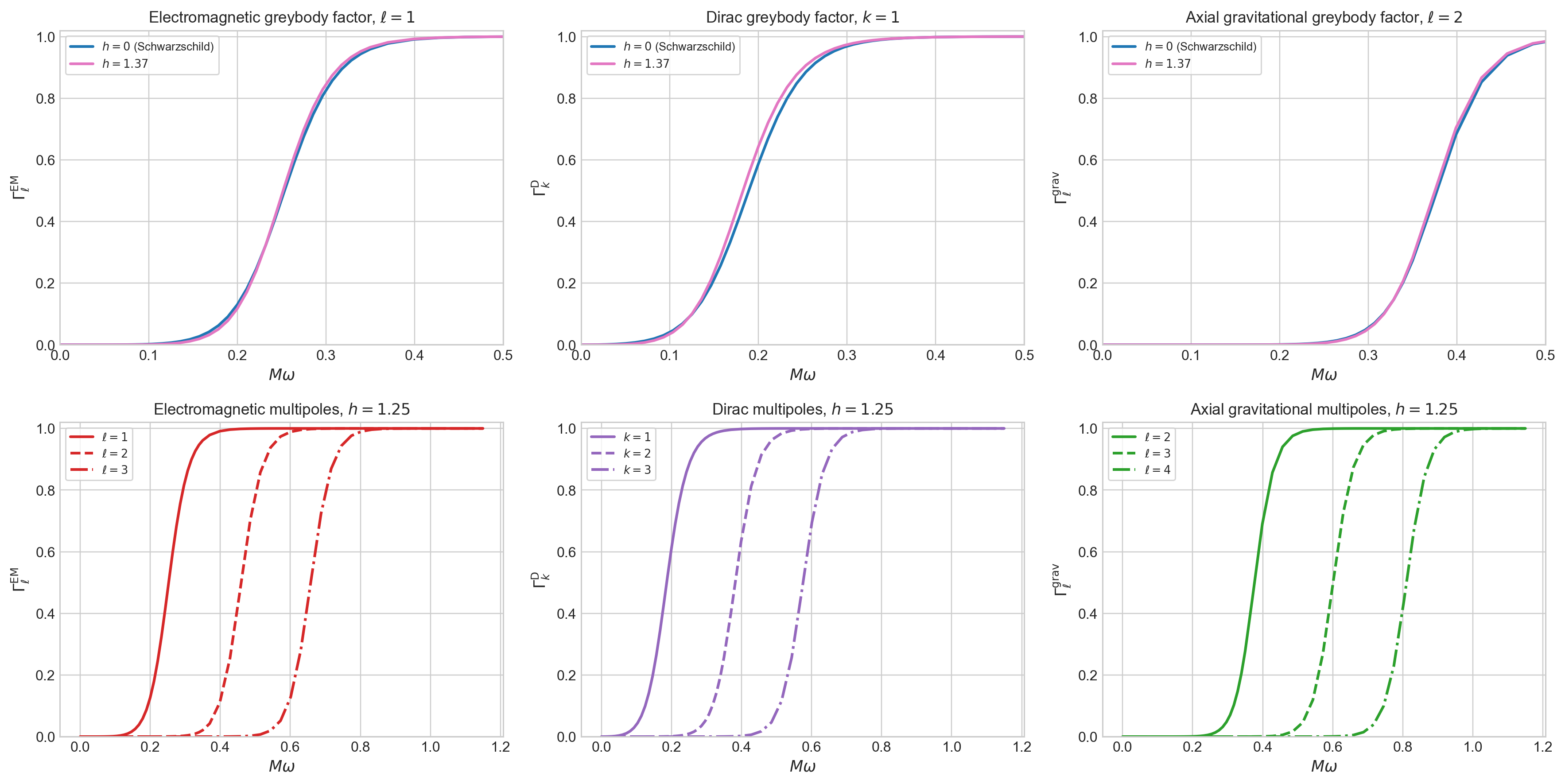}
\caption{Direct numerical greybody factors for electromagnetic and massless Dirac fields and for the gravitational axial-potential channel.  The upper panels compare the dominant modes for $h=0$ and the near-extremal value $h=1.37$.  The lower panels show the multipole hierarchy at $h=1.25$.  The right column gives the gravitational channel computed from Eq.~\eqref{eq:grav_potential}.}
\label{fig:gbf}
\end{figure*}

Figure~\ref{fig:gbf} shows the direct-integration greybody factors.  The electromagnetic threshold for $\ell=1$ is almost unchanged along the branch.  The Dirac threshold for $k=1$ shifts slightly to smaller $M\omega$ as the extremal endpoint is approached, but the shift is modest.  The gravitational axial-potential threshold for $\ell=2$ is also nearly fixed, moving only from $M\omega_{1/2}\simeq0.378$ in the Schwarzschild limit to $M\omega_{1/2}\simeq0.375$ at $h=1.37$.  Higher multipoles turn on at larger frequencies in the expected order.

Table~\ref{tab:half} gives half-transmission frequencies, defined by $\Gamma=1/2$, for the first three electromagnetic, Dirac and gravitational axial-potential modes.  The dominant electromagnetic half point changes from $M\omega\simeq0.253$ at Schwarzschild to $M\omega\simeq0.252$ at $h=1.37$.  The dominant Dirac half point changes from $M\omega\simeq0.189$ to $M\omega\simeq0.183$.  The dominant gravitational half point changes from $M\omega\simeq0.378$ to $M\omega\simeq0.375$.  Thus the Dymnikova deformation does not produce a large transparency enhancement at fixed ADM mass.

\begin{table*}[t]
\centering
\caption{Direct-integration half-transmission frequencies $M\omega_{1/2}$ for electromagnetic and Dirac modes and for the gravitational axial-potential modes.  The same gravitational greybody factors are used for the polar channels in the emission estimate through Eq.~\eqref{eq:axial_polar_assumption}.}
\label{tab:half}
\begingroup
\scriptsize
\setlength{\tabcolsep}{3pt}
\begin{ruledtabular}
\begin{tabular}{cccccccccc}
$h$ & EM $\ell=1$ & EM $\ell=2$ & EM $\ell=3$ & D $k=1$ & D $k=2$ & D $k=3$ & G $\ell=2$ & G $\ell=3$ & G $\ell=4$ \\
        0.0 & 0.253 & 0.461 & 0.659 & 0.189 & 0.383 & 0.576 & 0.378 & 0.602 & 0.811 \\
        0.8 & 0.253 & 0.461 & 0.659 & 0.189 & 0.383 & 0.576 & 0.378 & 0.602 & 0.811 \\
        1.1 & 0.253 & 0.460 & 0.659 & 0.189 & 0.383 & 0.576 & 0.377 & 0.602 & 0.811 \\
        1.25 & 0.253 & 0.460 & 0.658 & 0.187 & 0.382 & 0.576 & 0.377 & 0.601 & 0.811 \\
        1.32 & 0.252 & 0.459 & 0.658 & 0.185 & 0.381 & 0.575 & 0.376 & 0.600 & 0.810 \\
        1.36 & 0.252 & 0.459 & 0.657 & 0.183 & 0.380 & 0.574 & 0.375 & 0.600 & 0.810 \\
        1.37 & 0.252 & 0.458 & 0.657 & 0.183 & 0.380 & 0.574 & 0.375 & 0.600 & 0.810 \\
\end{tabular}
\end{ruledtabular}
\endgroup
\end{table*}

\begin{figure*}[t]
\centering
\includegraphics[width=0.90\textwidth]{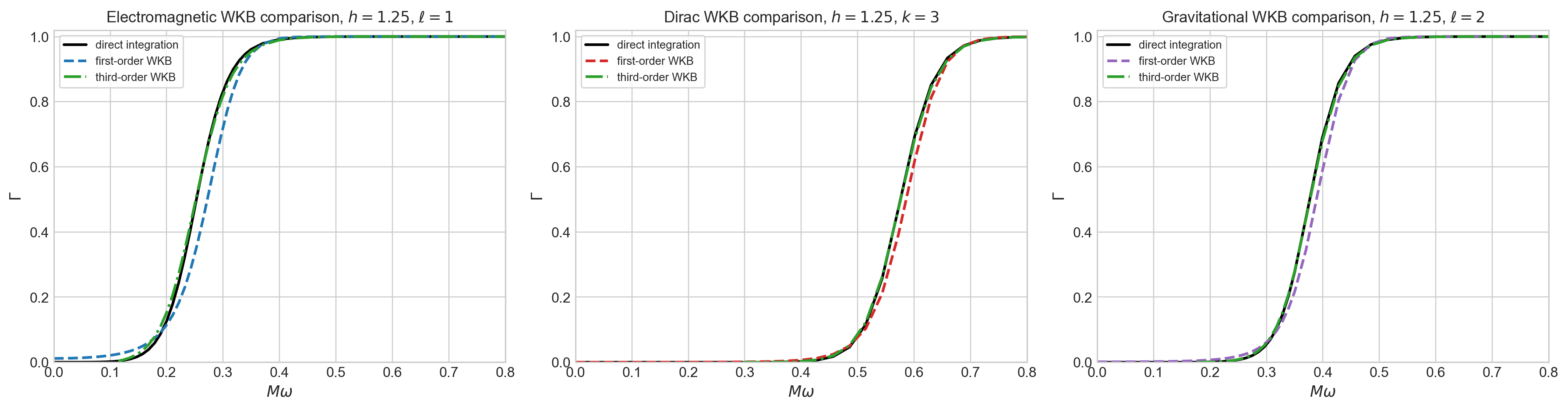}
\caption{Direct, first-order WKB and third-order WKB greybody factors at $h=1.25$.  The left panel shows the electromagnetic $\ell=1$ mode, the middle panel shows the Dirac $k=3$ mode, and the right panel shows the gravitational axial-potential $\ell=2$ mode.  As expected, WKB is most reliable near the barrier top and less reliable in the deep tunneling tail.}
\label{fig:wkb}
\end{figure*}

\begin{figure*}[t]
\centering
\includegraphics[width=0.95\textwidth]{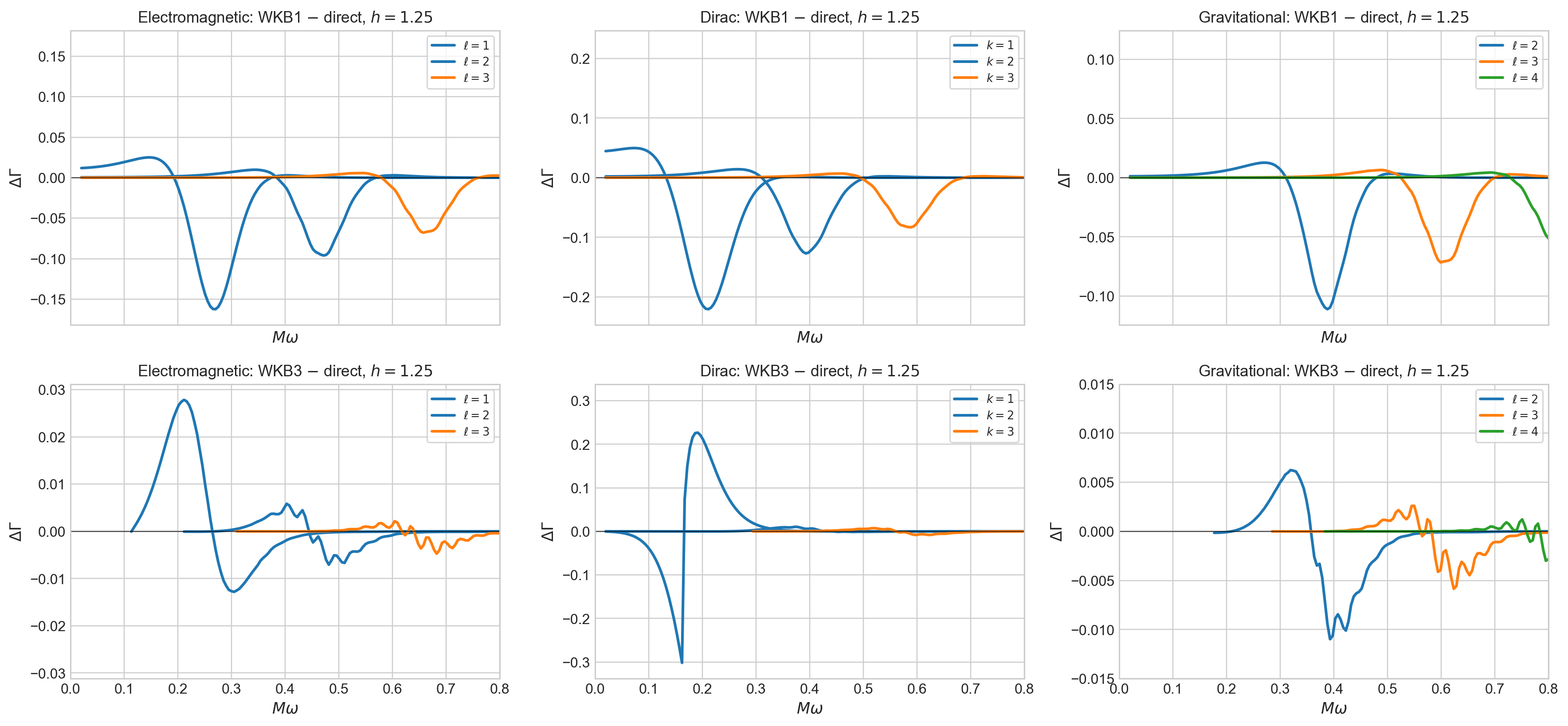}
\caption{Signed WKB residuals relative to the direct-integration greybody factors at $h=1.25$.  The plotted quantity is $\Delta\Gamma=\Gamma_{\rm WKB}-\Gamma_{\rm num}$ for the first three electromagnetic and Dirac multipoles and the first three gravitational axial-potential multipoles.  Curves are shown only where the continued WKB expression returns a finite real transmission probability.}
\label{fig:wkb_residuals}
\end{figure*}

The WKB comparison is shown in Fig.~\ref{fig:wkb}, while the signed residuals
\begin{equation*}
 \Delta\Gamma=\Gamma_{\rm WKB}-\Gamma_{\rm num}
\end{equation*}
are shown in Fig.~\ref{fig:wkb_residuals} for the first three electromagnetic and Dirac multipoles and for the first three gravitational axial-potential multipoles.  The residual plot gives a frequency-resolved check up to $M\omega=0.8$, instead of quoting only a few half-transmission points.  Third order substantially reduces the electromagnetic residuals and is also accurate for the gravitational modes and for Dirac $k=2$ and $k=3$ near the barrier transition, while the lowest Dirac mode retains the largest deviation.  This should be read as the expected accuracy of an asymptotic barrier-top approximation for low multipoles, not as an input to the spectra.  All emission rates use the direct-integration greybody factors.

\section{Energy-Emission Rates}

The emission rates in this section are computed for one static geometry at a time.  Equivalently, we use a quasi-canonical ensemble in which the Hawking temperature $T_H$ is fixed while a quantum is emitted; the temperature is updated only between successive adiabatic steps in the lifetime model.  This is the usual Page-type approximation and is justified when the black-hole mass is large compared with the typical emitted energy.

The energy spectrum per unit time and unit frequency is
\begin{equation}
\label{eq:emission_general}
 \frac{d^2E}{dt\,d\omega}=\frac{1}{2\pi}\sum_j
 \frac{N_j\,\omega\,\Gamma_j(\omega)}{\exp(\omega/T_H)\mp1},
\end{equation}
where the minus sign is for bosons and the plus sign is for fermions.  For one Maxwell field,
\begin{equation}
\label{eq:em_power}
 \frac{d^2E_\gamma}{dt\,d\omega}
 =\frac{1}{2\pi}\sum_{\ell=1}^{\infty}
 \frac{2(2\ell+1)\omega\Gamma_\ell^{\rm EM}}{\exp(\omega/T_H)-1}.
\end{equation}
For one two-helicity massless spin-$1/2$ species,
\begin{equation}
\label{eq:dirac_power}
 \frac{d^2E_D}{dt\,d\omega}
 =\frac{1}{2\pi}\sum_{k=1}^{\infty}
 \frac{2k\,\omega\Gamma_k^{\rm D}}{\exp(\omega/T_H)+1}.
\end{equation}
For one graviton field, using Eq.~\eqref{eq:axial_polar_assumption}, we take
\begin{equation}
\label{eq:grav_power}
  \frac{d^2E_g}{dt\,d\omega}
 =\frac{1}{2\pi}\sum_{\ell=2}^{\infty}
 \frac{2(2\ell+1)\,\omega\Gamma_{\ell}^{\rm grav,ax}}{\exp(\omega/T_H)-1}.
\end{equation}
The factor of two in Eq.~\eqref{eq:grav_power} counts the approximately degenerate axial and polar parity channels.  An axial-only luminosity would therefore be one half of the tabulated $P_g$.  The sums were truncated at $\ell_{\rm max}=k_{\rm max}=5$ for electromagnetic and Dirac fields and at $\ell_{\rm max}=6$ for the gravitational channel, which is sufficient at the temperatures considered here because the last included channels are negligible at the spectral peak.  We report five integrated quantities: $P_\gamma$ for one Maxwell field, $P_D$ for one two-helicity massless spin-$1/2$ species, $P_g$ for one graviton field in the axial--polar equality approximation, $P_\gamma+3P_D$ as a Page-like photon-plus-three-light-fermion proxy, and $12P_\gamma+24P_D$ as a massless Standard-Model-inspired gauge-plus-Dirac multiplicity count.
Here $j$ labels the partial wave and field species, $N_j$ is the corresponding degeneracy factor, and $\Gamma_j(\omega)$ is the transmission probability extracted from the radial scattering problem.  The multiplicity $P_\gamma+3P_D$ mimics one photon plus three light neutrino-like Dirac species, $P_g$ is kept separate because it relies on Eq.~\eqref{eq:axial_polar_assumption}, while $12P_\gamma+24P_D$ is only a massless high-temperature count of gauge and Dirac degrees of freedom; massive-particle thresholds are not switched on dynamically in the present calculation.

\begin{figure*}[t]
\centering
\includegraphics[width=\textwidth]{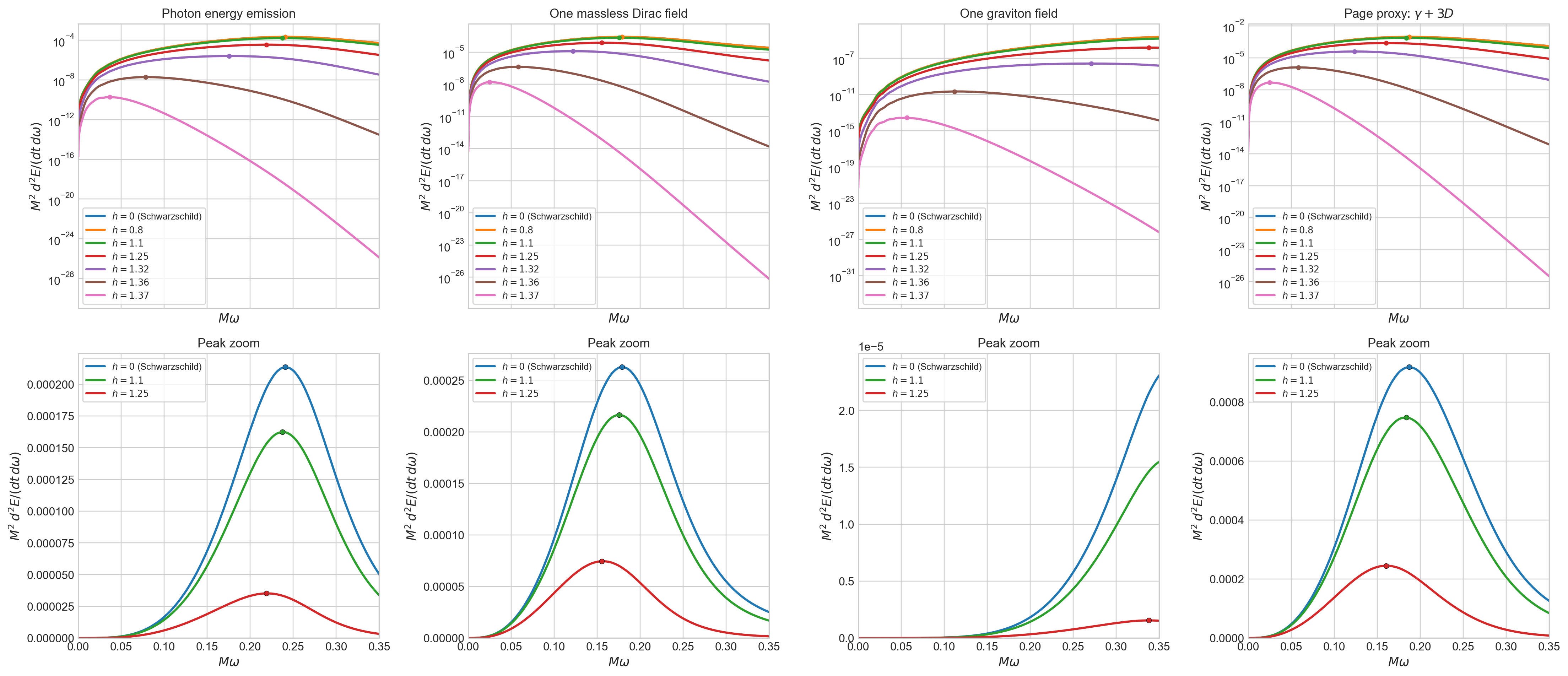}
\caption{Hawking energy-emission spectra for the Dymnikova black hole in units $M=1$.  The upper row shows logarithmic spectra for one photon field, one two-helicity massless spin-$1/2$ species, one graviton field under the axial--polar equality assumption and the Page proxy $P_\gamma+3P_D$.  The lower row zooms in on representative low-$h$ peaks.  The near-extremal spectra are strongly quenched by the falling Hawking temperature.}
\label{fig:emission}
\end{figure*}

Figure~\ref{fig:emission} shows the spectra.  The peak moves to lower frequency and its amplitude falls by many orders of magnitude as $h$ approaches $h_{\rm ext}$.  Since the greybody factors change only weakly, this collapse is driven mainly by the thermal factors in Eqs.~\eqref{eq:em_power}, \eqref{eq:dirac_power} and \eqref{eq:grav_power}.

The integrated powers are given in Table~\ref{tab:powers}.  The Schwarzschild row agrees with the expected Page-scale normalization for photons and massless fermions at fixed $M=1$.  For small $h$ the geometry is nearly indistinguishable from Schwarzschild outside the horizon.  Once $h$ reaches $1.25$, however, the Page proxy is already reduced to $2.32\times10^{-1}$ of its Schwarzschild value.  At $h=1.32$ it is $3.20\times10^{-2}$, at $h=1.36$ it is $6.14\times10^{-4}$, and at $h=1.37$ it is only $1.06\times10^{-5}$.  This occurs even though the Dirac threshold is slightly lower, because the exponential thermal suppression dominates the mild scattering change.

\begin{table*}[t]
\centering
\caption{Integrated powers in units $M=1$.  Here $P_\gamma$ is one Maxwell field, $P_D$ is one two-helicity massless spin-$1/2$ species, $P_g$ is one graviton field computed from the axial potential with the polar channel approximated by Eq.~\eqref{eq:axial_polar_assumption}, $P_\gamma+3P_D$ is the Page-like photon-plus-light-fermion proxy, and $12P_\gamma+24P_D$ is a massless Standard-Model-inspired multiplicity count.  The final column gives the Page-proxy ratio to the Schwarzschild value.}
\label{tab:powers}
\begingroup
\scriptsize
\setlength{\tabcolsep}{3pt}
\begin{ruledtabular}
\begin{tabular}{ccccccc}
$h$ & $P_\gamma$ & $P_D$ & $P_g$ & $P_\gamma+3P_D$ & $12P_\gamma+24P_D$ & Proxy ratio \\
        0.0 & $3.368\times10^{-5}$ & $4.093\times10^{-5}$ & $3.837\times10^{-6}$ & $1.565\times10^{-4}$ & $1.387\times10^{-3}$ & 1.000e+00 \\
        0.8 & $3.368\times10^{-5}$ & $4.093\times10^{-5}$ & $3.836\times10^{-6}$ & $1.565\times10^{-4}$ & $1.386\times10^{-3}$ & 1.000e+00 \\
        1.1 & $2.530\times10^{-5}$ & $3.296\times10^{-5}$ & $2.542\times10^{-6}$ & $1.242\times10^{-4}$ & $1.095\times10^{-3}$ & 7.936e-01 \\
        1.25 & $5.282\times10^{-6}$ & $1.034\times10^{-5}$ & $2.602\times10^{-7}$ & $3.629\times10^{-5}$ & $3.114\times10^{-4}$ & 2.319e-01 \\
        1.32 & $3.765\times10^{-7}$ & $1.543\times10^{-6}$ & $5.268\times10^{-9}$ & $5.006\times10^{-6}$ & $4.155\times10^{-5}$ & 3.199e-02 \\
        1.36 & $1.723\times10^{-9}$ & $3.144\times10^{-8}$ & $2.412\times10^{-12}$ & $9.604\times10^{-8}$ & $7.752\times10^{-7}$ & 6.138e-04 \\
        1.37 & $7.599\times10^{-12}$ & $5.488\times10^{-10}$ & $1.366\times10^{-15}$ & $1.654\times10^{-9}$ & $1.326\times10^{-8}$ & 1.057e-05 \\
\end{tabular}
\end{ruledtabular}
\endgroup
\end{table*}

At $h=0$ the geometry is exactly Schwarzschild, so the first row of Table~\ref{tab:powers} can be compared directly with Page's seminal calculation for an uncharged nonrotating black hole \cite{Page1976}.  Page's tabulated power coefficients in the same $G=c=\hbar=k_B=1$ normalization are $M^2P_\gamma=3.363\times10^{-5}$ for photons, $M^2P_{1/2}^{\rm Dirac}=8.167\times10^{-5}$ for a four-helicity Dirac field, and $M^2P_g\simeq3.84\times10^{-6}$ for gravitons.  Our Schwarzschild row gives $M^2P_\gamma=3.368\times10^{-5}$, $M^2P_D=4.093\times10^{-5}$ and $M^2P_g=3.837\times10^{-6}$.  Thus the photon and graviton channels agree with Page at the $0.2\%$ level or better, while $P_D$ is the two-helicity spin-$1/2$ normalization used in Eq.~\eqref{eq:dirac_power}; doubling it gives $2P_D=8.186\times10^{-5}$, which agrees with Page's four-helicity Dirac coefficient to $0.23\%$.  Combining the Page comparison set with four such two-helicity spin-$1/2$ species, $P_\gamma+4P_D+P_g$, gives $2.01\times10^{-4}$ and the familiar $81.4\%$, $16.7\%$ and $1.9\%$ split among spin-$1/2$, photon and graviton channels.

It is useful to spell out how the present calculation complements the recent analysis of Hawking radiation from renormalization-group-improved regular black holes by Konoplya \cite{Konoplya:2023bpf}, which includes a Dymnikova-type case.  In that paper the Dymnikova parameter is denoted by $l_{\rm cr}$; in our fixed-$M=1$ convention it corresponds to $h=(2l_{\rm cr}^2)^{1/3}$.  Table~\ref{tab:konoplya_comparison} therefore uses the same four Dymnikova points as Konoplya by recomputing, only for this comparison, the direct-scattering Page proxy $P_\gamma+3P_D$ at the mapped values of $h$.  These supplemental points are not added to the main benchmark tables. The particle content is also different.  Konoplya's luminosity is a Page-style total over the effectively massless Standard-Model degrees of freedom used in that work, including the fermionic, gauge-boson and scalar sectors with their multiplicity factors, but without a separate graviton channel in the quoted Dymnikova comparison.  The present column is deliberately narrower: it counts only one Maxwell field and three two-helicity massless Dirac species, $P_\gamma+3P_D$, while the axial--polar graviton estimate $P_g$ and the larger massless count $12P_\gamma+24P_D$ are left out of Table~\ref{tab:konoplya_comparison}. With the parameter mismatch removed, the remaining differences reflect the different particle-counting prescriptions and the use of WKB greybody factors in Ref.~\cite{Konoplya:2023bpf} versus direct scattering here; nevertheless, both calculations show an almost Schwarzschild luminosity for small cores and a rapid suppression toward the extremal endpoint.

\begin{table*}[t]
\centering
\begingroup
\caption{Numerical comparison with the Dymnikova-related Hawking-radiation analysis of Konoplya \cite{Konoplya:2023bpf} at the same parameter values.  The Konoplya column gives the total luminosity normalized to its Schwarzschild value.  The present-work columns give supplemental direct-scattering calculations at $h=(2l_{\rm cr}^2)^{1/3}$, using the Page proxy $P_\gamma+3P_D$ normalized to the Schwarzschild row of Table~\ref{tab:powers}.}
\label{tab:konoplya_comparison}
\scriptsize
\setlength{\tabcolsep}{3pt}
\begin{ruledtabular}
\begin{tabular}{cccccp{0.24\textwidth}}
Konoplya $l_{\rm cr}$ & mapped $h$ used here & Konoplya rate ratio & present $P_\gamma+3P_D$ & present proxy ratio & Interpretation \\
0.5 & 0.7937 & 0.9997 & $1.565\times10^{-4}$ & 0.99997 & Schwarzschild-like regime \\
0.8 & 1.0858 & 0.9443 & $1.298\times10^{-4}$ & 0.8295 & mild suppression; method and multiplicity differences visible \\
1.0 & 1.2599 & 0.4430 & $3.049\times10^{-5}$ & 0.1949 & onset of strong cooling \\
1.1 & 1.3426 & 0.0023 & $1.102\times10^{-6}$ & $7.04\times10^{-3}$ & near-remnant quenching \\
\end{tabular}
\end{ruledtabular}
\endgroup
\end{table*}

The gravitational contribution is numerically subdominant in this model: $P_g/(P_\gamma+3P_D)=2.45\times10^{-2}$ in the Schwarzschild row, $7.17\times10^{-3}$ at $h=1.25$, and $8.26\times10^{-7}$ at $h=1.37$.  These numbers already include the approximate polar contribution; without Eq.~\eqref{eq:axial_polar_assumption} they would be smaller by a factor of two.  The composition of the massless flux also changes.  For one Maxwell field and one two-helicity spin-$1/2$ species, $P_\gamma/P_D\simeq0.823$ at $h=0$, $0.511$ at $h=1.25$, $0.244$ at $h=1.32$, $5.48\times10^{-2}$ at $h=1.36$, and $1.38\times10^{-2}$ at $h=1.37$.  Thus the photon channel is quenched more strongly than the Dirac channel.  In the simple massless Standard-Model-inspired count, the near-remnant emission becomes increasingly fermion dominated.

\section{Adiabatic Evaporation Time}
\label{sec:lifetime}

The fixed-mass sequence above can be converted into a simple evaporation estimate by holding the Dymnikova core scale $r_0$ fixed while the ADM mass decreases.  This is an adiabatic model rather than a full back-reacting solution, but it captures the standard regular-black-hole expectation that a zero-temperature endpoint delays the final stage of evaporation \cite{Hayward2006}.  Thus the geometry is treated as a succession of equilibrium black holes: during one emission event the parameters are frozen, and after the emitted energy is removed from $M$ the system is moved to the next nearby member of the family.  Restoring $M$ in Eq.~\eqref{eq:dymnikova_original} and measuring radii in units of $M$ gives
\begin{equation}
\label{eq:h_running}
 h(M)^3=\frac{2r_0^2}{M^2},
 \qquad
 \frac{M}{r_0}=\sqrt{2}\,h^{-3/2}.
\end{equation}
Thus evaporation drives the system toward larger $h$, and the extremal remnant mass is
\begin{equation}
\label{eq:remnant_mass}
 M_{\rm ext}=\sqrt{2}\,r_0 h_{\rm ext}^{-3/2}.
\end{equation}
If ${\cal P}(h)=P_\gamma+3P_D$ denotes the Page-proxy luminosity tabulated at $M=1$, dimensional scaling gives $P(M,h)={\cal P}(h)/M^2$.  The symbol ${\cal P}$ is therefore dimensionless in the $M=1$ tables, while the physical luminosity has dimension $M^{-2}$.  The mass-loss equation therefore becomes
\begin{equation}
\label{eq:lifetime_integral}
\begin{aligned}
 \frac{dt}{r_0^3}
 &=3\sqrt{2}\,\frac{h^{-11/2}}{{\cal P}(h)}\,dh,\\
 \Delta t(h_i\to h_f)
 &=3\sqrt{2}\,r_0^3
 \int_{h_i}^{h_f}\frac{h^{-11/2}}{{\cal P}(h)}\,dh .
\end{aligned}
\end{equation}
For $h\le1.37$ we interpolate $\log[{\cal P}(h)]$ with a monotone cubic spline; for the small interval from $1.37$ to $h_f=0.999h_{\rm ext}$ we use a low-temperature power-law extrapolation in $M T_H$, fitted to the last directly computed points.  This tail affects the precise cutoff time but not the qualitative conclusion, because $T_H\to0$ at the endpoint.

\begin{table*}[t]
\centering
\caption{Adiabatic fixed-core evaporation estimates for the Page-proxy luminosity.  The Dymnikova scale $r_0$ is held fixed, so $h$ increases as the ADM mass decreases.  The time column gives the interval from the listed $h_i$ to $h_f=0.999h_{\rm ext}$.  The final column divides by the Schwarzschild lifetime of a black hole with the same initial mass and the same Page-proxy normalization.}
\label{tab:lifetime}
\begin{ruledtabular}
\begin{tabular}{cccccc}
$h_i$ & $M_i/M_{\rm ext}$ & $T_H r_0$ & $(P_\gamma+3P_D)r_0^2$ & $\Delta t/r_0^3$ & $\Delta t/t_{\rm Schw}$ \\
        0.8 & 2.249 & $0.0201$ & $4.006\times10^{-5}$ & $4.358\times10^{6}$ & 2.650e+02 \\
        1.1 & 1.395 & $0.031$ & $8.265\times10^{-5}$ & $4.345\times10^{6}$ & 1.107e+03 \\
        1.25 & 1.151 & $0.03$ & $3.544\times10^{-5}$ & $4.341\times10^{6}$ & 1.967e+03 \\
        1.32 & 1.061 & $0.0233$ & $5.757\times10^{-6}$ & $4.336\times10^{6}$ & 2.510e+03 \\
        1.36 & 1.015 & $0.0131$ & $1.208\times10^{-7}$ & $4.278\times10^{6}$ & 2.833e+03 \\
        1.37 & 1.004 & $6.785\times10^{-3}$ & $2.126\times10^{-9}$ & $3.319\times10^{6}$ & 2.272e+03 \\
\end{tabular}
\end{ruledtabular}
\end{table*}

Table~\ref{tab:lifetime} shows that most of the remaining time is spent close to the cold endpoint.  Starting from $h_i=0.8$, where the fixed-mass luminosity is essentially Schwarzschild-like, the time to reach $0.999h_{\rm ext}$ is already $4.36\times10^6r_0^3$, about $2.65\times10^2$ times the Schwarzschild lifetime for the same initial mass.  Starting closer to the remnant increases this ratio because the luminosity is more strongly suppressed over the whole interval.  This is a qualitative difference from Schwarzschild evaporation: instead of the temperature increasing without bound as $M$ decreases, the Dymnikova branch cools toward a finite-mass, zero-temperature state.  The exact extremal point is reached only asymptotically within this extrapolated model, so the finite numbers in Table~\ref{tab:lifetime} should be read as cutoff times to a specified near-extremal state rather than as complete evaporation times.

\section{Conclusion}

We have computed electromagnetic, massless Dirac and gravitational greybody factors and Hawking energy-emission rates for the Dymnikova regular black hole.  In units $M=1$, the metric is controlled by one length parameter $h=(2r_0^2)^{1/3}$.  The double-horizon conditions give
\begin{equation*}
\begin{aligned}
 h_{\rm ext}&=1.373256824469,\\
 r_{\rm ext}&=1.702001406975,\\
 r_{0,{\rm ext}}&=1.137922411780.
\end{aligned}
\end{equation*}
The branch therefore ends in a zero-temperature extremal remnant.  All numerical results are obtained in the fixed-background, fixed-temperature ensemble described above, and the evaporation curve is an adiabatic reconstruction from these equilibrium data.

Direct integration shows that the electromagnetic, Dirac and gravitational axial-potential greybody thresholds are only mildly affected by the Dymnikova deformation at fixed ADM mass.  The gravitational emission quoted here additionally assumes approximate equality of the polar and axial greybody factors.  Third-order WKB gives small frequency-resolved residuals for the electromagnetic, gravitational and higher-Dirac barrier transitions, while the lowest Dirac mode demonstrates the expected limitation of a low-multipole WKB approximation.  The spectra and integrated powers are therefore governed primarily by thermodynamics rather than by a large change in exterior transparency.

The resulting Hawking luminosity is strongly quenched near the endpoint.  The gravitational channel follows the same thermal quenching and is smaller than the photon-plus-light-fermion proxy throughout the benchmark sequence.  Relative to its Schwarzschild value, the Page-like photon-plus-three-Dirac proxy falls to $2.32\times10^{-1}$ at $h=1.25$, $3.20\times10^{-2}$ at $h=1.32$, $6.14\times10^{-4}$ at $h=1.36$, and $1.06\times10^{-5}$ at $h=1.37$.  The photon sector decreases faster than the Dirac sector, so the massless flux becomes progressively fermion dominated as the cold remnant is approached.  The large suppression is not mainly a greybody-factor effect.  It is a thermodynamic effect caused by the vanishing surface gravity, with the greybody factors providing spin-dependent corrections to the residual flux composition.

The evaporation history is therefore qualitatively different from that of a Schwarzschild black hole.  In the Schwarzschild case the semiclassical temperature rises as the mass decreases, leading to an accelerating evaporation rate within the fixed-background approximation.  In the fixed-core Dymnikova model studied here, the same decrease of the ADM mass instead drives the system toward larger $h$ and lower $T_H$, so the luminosity collapses and the approach to the extremal configuration is delayed indefinitely in the zero-temperature limit.  This conclusion is robust at the level of the present test-field calculation, although the precise late-time cutoff depends on the assumed low-temperature extrapolation and on possible backreaction beyond the adiabatic model.

The physical mass range should also be interpreted with care.  The computation is not intended to describe the final Planckian regime if $r_0$ is microscopic and $M_{\rm ext}=O(M_{\rm Pl})$.  It instead gives a controlled semiclassical description whenever the instantaneous mass is large compared with both the Planck mass and the energy of an emitted quantum, or whenever the effective regularization scale is large enough that the remnant mass is still semiclassical.  Under these conditions the emitted Standard-Model particles are light compared with the black hole and the neglect of recoil and single-quantum backreaction is self-consistent; if $T_H$ drops below a particle rest mass, that species should be removed by a massive-threshold treatment not included here.

It is worth emphasizing that the grey-body factors computed in the present work have a significance extending well beyond their conventional role in determining the spectrum and intensity of Hawking radiation. In recent years, it has been demonstrated that the transmission and reflection properties of black-hole effective potentials encode detailed information about the quasinormal spectrum, leading to a nontrivial correspondence between grey-body factors and quasinormal modes \cite{Konoplya:2024vuj,Konoplya:2024lir,Dubinsky:2024vbn,Lutfuoglu:2025mqa,Skvortsova:2024msa,Bolokhov:2024otn,Malik:2024cgb}. Within this framework, the frequency dependence of the grey-body factors can be employed to infer the characteristics of damped oscillatory modes governing the ringdown stage of black-hole perturbations. Consequently, the grey-body factors obtained here provide an additional probe of the spacetime geometry, complementing direct quasinormal-mode calculations and potentially allowing one to extract spectral information from scattering data alone.

Future extensions should include massive Standard-Model thresholds, a full polar gravitational perturbation calculation beyond the axial--polar equality approximation, rotating Dymnikova-type geometries, and a fully dynamical evaporation model in which the core parameter and the ADM mass evolve together rather than being sampled as a fixed-mass family.  Such extensions are needed before assigning a unique phenomenological lifetime to a specific microscopic model, but the present results already show the main mechanism: a regular de Sitter core with an extremal endpoint can convert the usual Schwarzschild runaway into a long, cold, remnant-dominated evaporation stage.

\bibliographystyle{apsrev4-1}
\bibliography{references}

\end{document}